\documentclass[5p,twocolumn,times]{elsarticle}
\journal{Physics Letters B}

\usepackage{graphicx}
\usepackage{amsmath}
\usepackage{amssymb}

\newcommand{\ba}{\begin{eqnarray}}
\newcommand{\ea}{\end{eqnarray}}
\newcommand{\bmath}{\begin{mathletters}}
\newcommand{\emath}{\end{mathletters}}
\newcommand{\ban}{\begin{eqnarray*}}
\newcommand{\ean}{\end{eqnarray*}}

\newcommand{\bsub}{\begin{subequations}}
\newcommand{\esub}{\end{subequations}}

\begin{document}


\title
{Effect of a fermion on quantum phase transitions in bosonic systems}

\author[yale]{F. Iachello}
\ead{francesco.iachello@yale.edu}

\author[huji]{A. Leviatan\corref{cor1}}
\ead{ami@phys.huji.ac.il}

\author[demokritos]{D. Petrellis}
\ead{petrellis@inp.demokritos.gr}

\cortext[cor1]{Corresponding author}

\address[yale]{Center for Theoretical Physics, Sloane Physics Laboratory, 
Yale University, New Haven, Connecticut~06520-8120, USA}
\address[huji]{Racah Institute of Physics, The Hebrew University, 
Jerusalem 91904, Israel}
\address[demokritos]
{Institute of Nuclear Physics, N.C.S.R. 
``Demokritos'', GR-15310 Aghia Paraskevi, Attiki, Greece}

\begin{abstract}
The effect of a fermion with angular momentum $j$ on quantum phase 
transitions of a ($s,d$) bosonic system is investigated. It is shown that 
the presence of a fermion strongly modifies 
the critical value at which the transition occurs, and its nature, even for 
small and moderate values of the coupling constant. The analogy with a 
bosonic system in an external field is mentioned. Experimental evidence for 
precursors of quantum phase transitions in bosonic systems plus a fermion 
(odd-even nuclei) is presented.
\end{abstract}



\maketitle

Quantum phase transitions (QPT) are qualitative changes in the ground state
properties of a physical system induced by a change in one or more
parameters in the quantum Hamiltonian describing the system. Originally
introduced in the 1970's~\cite{hertz,gilmore1}, they have been the subject 
in recent years of many investigations and have found a variety of
applications in many areas of physics and chemistry~\cite{vojta,carr}. 
One of these applications is to atomic nuclei, where QPTs have been
extensively investigated (for a review, see~\cite{iac3,casten11,iac11}) 
within the framework of the Interacting Boson Model (IBM), a model of 
even-even nuclei in terms of correlated pairs of valence nucleons 
with angular momentum $J=0,2$ treated as bosons ($s,d$)~\cite{iac1}. 
For this case also finite size effects~\cite{levgin03,zamfir,lev0506} 
and scaling behavior~\cite{rowe04,dusuel,williams} have been
investigated, both analytically and numerically, 
showing that precursors of QPT can be seen even for relatively small
values of~$N$. \ QPTs have also been extended to excited states quantum
phase transitions, that is qualitative changes in the properties of
the system as a function of the excitation energy~\cite{caprio}. In this
letter we present results of an investigation of the effect of a fermion on
QPTs in bosonic systems. We do this in atomic nuclei by making use of the
Interacting Boson Fermion Model (IBFM), a model of odd-even nuclei in terms
of correlated pairs with angular momentum $J=0,2$ ($s,d$ bosons) and
unpaired particles with angular momentum $J=j$ ($j$ fermions)~\cite{iac2}.
As an illustration we take $j=11/2$. We note, however, that our method of
analysis can also be used for systems with other values of the fermion, $j$,
and boson, $J$, angular momenta, for example the spin-boson systems
discussed in~\cite{lehur}, the simplest case of which is a fermion with 
$j=1/2$ (\textit{i.e.}, a single spin) in a bath of harmonic oscillator 
one-dimensional bosons of interest in dissipation and light phenomena. 
QPTs in IBFM for selected orbits have been investigated by 
Alonso \textit{et al.}~\cite{alonso1,boyukata}. 
Here we focus on the effect of a fermionic impurity on QPTs in bosonic 
systems. Our main results are that, 
(1)~the presence of a single fermion greatly influences the 
location and nature of the phase transition, the fermion acting either as 
a catalyst or a retarder of the QPT, and (2)~there is experimental evidence 
for quantum phase transitions in odd-even nuclei (bosonic systems plus 
a single fermion).

Within the context of the geometric collective model of nuclei, the effect 
of an odd particle on collective properties was investigated years ago in 
core-particle models. However, both our results are novel, 
since (i)~in the deformed phase the effect of the fermion is of 
order $1/N$ and thus vanishes in the geometric limit $N\rightarrow\infty$. 
However, we explicitly show that the effect is large in transitional nuclei 
even for small and moderate values of the coupling strength of the fermion 
to the bosons; (ii)~the experimental evidence for QPTs in odd-nuclei has not 
been presented earlier and we show one of the key signatures of QPTs in 
nuclei, the two-neutron separation energies. This quantity is discontinuous 
for a first order phase transition at the transition point (or it has 
a sudden jump for a finite system). As shown below, this jump is observed 
experimentally.

To prove our point, we consider the Hamiltonian of a system of $N$ ($s,d$) 
bosons coupled (with a quadrupole interaction) to a single fermion with 
angular momentum $j$~\cite{iac2}
\begin{equation}
H=H_{B}+H_{F}+V_{BF}~,
\label{Eq:hBF}
\end{equation}
with
\ba
H_{B} &=& 
\varepsilon _{0}\left[ \left( 1-\xi \right) \hat{n}_{d}-\frac{\xi }{
4N}\hat{Q}^{\,\chi }\cdot \hat{Q}^{\,\chi }\right]~,  
\nonumber \\
H_{F} &=& 
\varepsilon _{j}~,  
\nonumber \\
V_{BF} &=& 
\Gamma \hat{Q}^{\,\chi }\cdot \hat{q}~.
\label{Eq:hBhFvBF}
\ea
Here $\hat{n}_{d}= d^{\dag }\cdot \tilde{d}$ is the number operator 
for d-bosons, 
$\hat{Q}^{\,\chi }=( d^{\dag }\times s+s^{\dag }\times \tilde{d})^{(2)}
+\chi ( d^{\dag }\times \tilde{d})^{(2)}$ and 
$\hat{q}=( a_{j}^{\dag }\times \tilde{a}_{j})^{(2)}$, 
are quadrupole operators of bosons and fermion respectively, 
$\varepsilon _{0}$ is the scale of the boson energy, $\varepsilon _{j}$ 
is the energy of the single fermion and $\Gamma$ the strength of the 
quadrupole Bose-Fermi interaction.
The dot and cross indicate scalar and tensor products and the adjoint
operators for bosons and fermions are $\tilde{d}_{\mu }=\left( -\right)
^{\mu }d_{-\mu }$ and $\tilde{a}_{j,m}=\left( -\right) ^{j-m}a_{j,-m}$. QPTs
of the purely bosonic part of the Hamiltonian $H_{B}$ have been extensively
investigated~\cite{diep,feng}. There are two control parameters $\xi 
$ and $\chi $. For fixed $\chi $, as one varies $\xi $, $0\leq \xi \leq 1$,
the bosonic system undergoes a QPT. The phase transition is first order for 
$\chi \neq 0$ and becomes second order at $\chi =0$. No phase transition
occurs as a function of $\chi $. In this article, we take 
$\chi =-\frac{\sqrt{7}}{2}$, in which case the two ``phases" of the system 
have U(5) symmetry ($\xi =0$) and SU(3) symmetry ($\xi =1$)~\cite{iac1}. 
The critical point, separating the spherical [U(5)] 
and axially-deformed [SU(3)] phases, 
occurs at $\xi_{c}\cong 1/2$. 

A complete study of the properties of quantum phase transitions 
necessitates both a classical and a quantal analysis, 
and a consideration of other couplings of 
fermions to bosons in addition to quadrupole coupling~\cite{petrellis}. 
In order to emphasize the main features of 
the results, we report here only the
classical analysis. This amounts to constructing the combined Bose-Fermi
potential energy surface (Landau potential) and minimizing it with respect
to the classical variables. To this end, we introduce a 
boson condensate~\cite{iac1} 
\begin{equation}
\left\vert N;\beta ,\gamma \right\rangle =\frac{1}{\sqrt{N!}}\left[
b_{c}^{\dag }\left( \beta ,\gamma \right) \right] ^{N}
\left\vert 0\right\rangle~,
\label{Eq:cond}
\end{equation}
$b_{c}^{\dag} = (1+\beta ^{2})^{-1/2}[ \beta\cos \gamma d_{0}^{\dag } 
+ \beta \sin \gamma (d_{2}^{\dag }+d_{-2}^{\dag })/\sqrt{2} +s^{\dag }]$, 
in terms of the classical variables $\beta ,\gamma $. The expectation 
value of $H_{B}$ of Eq.~(\ref{Eq:hBhFvBF}) in the condensate is~\cite{zamfir} 
\ba
&&E_{B}\left (N;\beta ,\gamma \right ) 
=\left\langle N;\beta ,\gamma \left\vert
H_{B}\right\vert N;\beta ,\gamma \right\rangle   
\nonumber \\
&&=\varepsilon _{0} N\left \{\frac{\beta ^{2}}{1+\beta ^{2}}
\left [1-\xi - ( \chi ^{2}+1 ) \frac{\xi }{4N} \right ] -
\frac{5\xi }{4N(1+\beta ^{2})}
\right.  
\nonumber \\
&&
\left. 
-\frac{\xi }{(1+\beta ^{2})^{2}}\frac{N-1}{N}
\left[ \beta ^{2}-\sqrt{\frac{2}{7}}\chi \beta ^{3}\cos 3\gamma 
+\frac{1}{14}\chi ^{2}\beta ^{4}\right] \right\}.
\nonumber\\
\label{Eq:EB}
\ea
We then evaluate the expectation value of $H_{F}$ and $V_{BF}$ in the
condensate thus obtaining a fermion Hamiltonian 
\ba
{\cal H}(N;\beta ,\gamma ) &=& E_{B}\left (N;\beta ,\gamma \right )  
\nonumber \\
&& +\sum_{m_{1},m_{2}}\left[\,\varepsilon _{j}\,\delta_{m_{1},m_{2}}
+ g_{m_{1},m_{2}}(N;\beta ,\gamma )\, \right]  
\nonumber \\
&&\qquad\times \left( \frac{a_{j,m_{1}}^{\dag }a_{j,m_{2}}
+ a_{j,m_{2}}^{\dag}a_{j,m_{1}}}{1+\delta _{m_{1},m_{2}}}\right).
\label{Eq:Hdef}
\ea 
The matrix $g_{m_{1},m_{2}}(N;\beta ,\gamma )$ is a real, 
symmetric matrix 
\ba
&&
g_{m_{1},m_{2}}(N;\beta ,\gamma ) =
N\Gamma \left( \frac{\beta}{1+\beta ^{2}}\right) \left( -\right) ^{j+m_{2}}  
\nonumber \\
&&
\qquad
\times \left\{ \left[ 2\cos \gamma 
-\chi \sqrt{\frac{2}{7}}\beta \cos 2\gamma \right]
C_{j,m_{1};j,-m_{2}}^{2,0}
\right. 
\nonumber\\ 
&&
\qquad
\left. +\left[ \sqrt{2}\sin \gamma 
+\chi \sqrt{\frac{1}{7}}\beta \sin 2\gamma \right]
C_{j,m_{1};j,-m_{2}}^{2,2}\right\},\;\;
\label{Eq:gmm}
\ea 
where $C_{j,m_{1};j,-m_{2}}^{2,m}$ 
denotes a Clebsch Gordan coefficient. 
The eigenvalues $e_{i}$ and eigenvectors $\psi _{i}$ of the matrix $g$ are 
the single-particle energies and wave functions of the fermion in the
deformed ($\beta ,\gamma $) field generated by the bosons. 
For $\gamma= 0^{\circ}$ (field with axial symmetry), 
$\chi = 0$ and $\beta$ small ({\it i.e.}, neglecting $\beta^2$) 
they were obtained years ago by Nilsson~\cite{Nilsson}. 
For $\gamma\neq 0^{\circ}$; $\chi  = 0$ and  $\beta$ small,
they were investigated by Meyer-ter-Vehn~\cite{Meyer}. 
We have solved the problem 
in its generality and details are given in~\cite{petrellis}. Here, we consider, 
for simplicity, the case $\gamma = 0^{\circ}$ for which the eigenvalues 
are given in explicit analytic form~\cite{leviatan}
\ba
\lambda _{K}\left( N;\beta ;\chi ;\Gamma \right) &=&
-N\Gamma 
\left( 
\frac{\beta }{1+\beta ^{2}}
\right) 
\sqrt{5}\left( 2-\beta \chi \sqrt{\frac{2}{7}}\right)
\nonumber \\ 
&&\;\;\;
\times P_{j}[3K^{2}-j(j+1)]\frac{}{}~,
\label{Eq:lamK}
\ea 
where $P_{j}=\left[ (2j-1)j(2j+1)(j+1)(2j+3)\right] ^{-1/2}$. 
\begin{figure}[t]
\center{{\includegraphics[height=140mm]{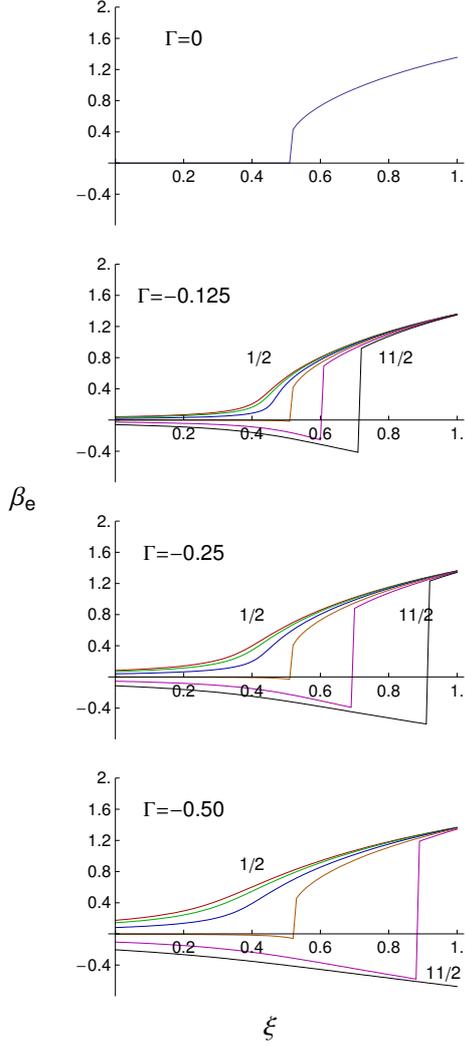}}} 
\caption{
(color online). 
Equilibrium values of the total boson plus fermion energy 
(classical order parameter) as a
function of the control parameter $\xi $ of the bosonic phase transition for
various values of the coupling constant $\Gamma $ in units of 
$\varepsilon _{0}$ and $N=10$. 
The curves for $\Gamma\neq 0$ correspond, from left to right, to states 
with $K=1/2,3/2,5/2,7/2,9/2,11/2$.
The value $\Gamma =0$ gives the purely bosonic case.}
\label{fig1}
\end{figure}
The quantum number $K=j,j-1,j-2,...,\frac{1}{2}$ 
has the physical meaning of 
the projection of the angular momentum on the intrinsic $z$ axis of the
condensate.

Once the eigenvalues have been obtained, one can calculate the total energy
functional (Landau potential for the combined Bose-Fermi system)
\ba
E_{i}\left (N;\beta ,\gamma ;\xi ,\chi ;\Gamma \right ) &=& 
E_{B}\left( N;\beta ,\gamma ;\xi,\chi \right)
\nonumber\\
&&
+\, \varepsilon_{j} 
+ e_{i}\left (N;\beta ,\gamma ;\chi ;\Gamma \right )~.
\label{Eq:Ei}
\ea 
This expression is the algebraic analog of the total potential 
energy surface, obtained in the macroscopic-microscopic Strutinsky 
procedure~\cite{heyde83}. 
Minimization of $E_{i}$ with respect to $\beta$ and $\gamma $ gives the
equilibrium values $\beta_{e},\gamma_{e}$ (the classical order
parameters) for each state. In the simple case of
$\gamma =0^{\circ },\,\chi =-\frac{\sqrt{7}}{2},\,\varepsilon _{j}=0$ 
the total energy functional becomes 
\ba
E_{K}\left (N;\beta ;\xi ;\Gamma \gamma \right ) &=& 
E_{B}\left (N;\beta ,0;\xi ,-\frac{\sqrt{7}}{2}\right )
\nonumber\\
&&+\, \lambda _{K}(N;\beta; -\frac{\sqrt{7}}{2};\Gamma )~.
\label{Eq:EK}
\ea
Minimization ($\frac{\partial E_{K}}{\partial \beta }=0$) gives the
equilibrium $\beta _{e}$ values shown in Fig.~\ref{fig1} 
as a function of the control
parameter $\xi $ of the bosonic phase transition. 
\begin{figure}
\center{{\includegraphics[height=50mm]{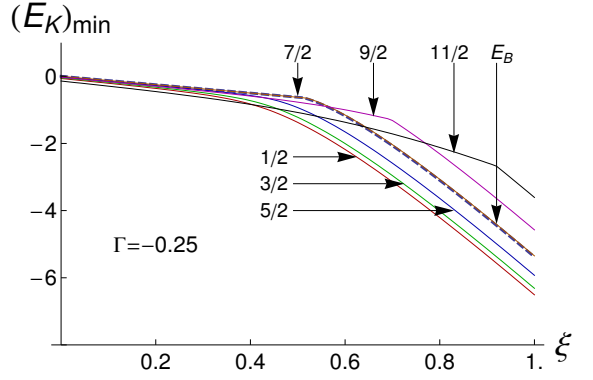}}}
\caption{
(color online). 
The total energy at the equilibrium value, 
$(E_{K})_{\min }$ Eq.~(\ref{Eq:EK}), as a function of the 
control parameter $\xi$ 
for an intermediate value of the strength of the Bose-Fermi 
interaction, $\Gamma =-0.25$, with $\varepsilon _{0}=1$ and $N=10$.
The dashed curve, labeled by $E_B$, is the corresponding energy of 
the purely bosonic system.}
\label{fig2}
\end{figure}

By comparing the 
top part of this figure (purely bosonic system) with the bottom part, one 
can see that the effect of the fermionic impurity is to wash out the phase 
transition for states with $K = 1/2,\, 3/2,\, 5/2$ and to enhance it for 
states with $K = 7/2,\, 9/2,\, 11/2$. In other words, the fermion acts as 
a catalyst for some states and as a retarder for others. 
Also, when the coupling strength becomes very large, the minima for some 
large $K$, in the figure $K = 11/2$, shift to negative values (oblate 
deformation). In addition to this result, also known qualitatively from 
particle-core models, the effect of the fermion is to move the location of 
the critical point even for small and moderate values of $\Gamma$. 
Physical values of $\Gamma$ in the $_{61}$Pm, $_{63}$Eu, $_{65}$Tb nuclei,
where the phase transition occurs, are $\Gamma\cong -0.125$~\cite{Scholten}.

It is interesting to note that the effect of the fermionic impurity is
similar to the effect of an external field with linear coupling on a
thermodynamic phase transition investigated by Landau and Lifshitz years 
ago~\cite[p.456]{landau}. They considered the potential 
\begin{equation}
\Phi (\eta )=\Phi _{0}+A\eta ^{2}+B\eta ^{4}+\alpha\, \eta~,
\label{Eq:Landau}
\end{equation}
where $\alpha $ is the strength of the coupling to the external field and 
$\eta $ a classical variable (order parameter). (The bosonic part of this
potential $A\eta ^{2}+B\eta ^{4}$ has only a second order transition). 
After a projective transformation 
$\frac{\beta ^{2}}{1+\beta^{2}}=\eta ^{2}$ 
which does not change the nature of the phase transition
and some rearrangement, the IBFM potential can be written
for small $\eta$, in the form
\begin{equation}
E_{K} (\eta) = 
\Delta_{0}+A\eta ^{2}+C\eta ^{3}+B\eta^{4}
+\alpha _{K}\left( 2\eta + \frac{\eta ^{2}}{\sqrt{2}}\right)~,\;
\label{Eq:EKeta}
\end{equation}
where $\alpha _{K}$ is the strength of the coupling for each $K$ value. By
comparing the couplings in Eqs.~(10) and (11), one can see that the IBFM 
$E_{K}\left (\eta \right )$ is more general than $\Phi (\eta )$ since it 
has a linear and
quadratic coupling, but for small~$\eta $ (note that 
$\left\vert \eta \right\vert \leq 1$), the quadratic term is negligible
and the two expressions become identical. 
Also, in the IBM the bosonic part has a cubic term
leading to the possibility of a first order transition. 
The analogy with bosonic systems in an external field also suggests that 
our results apply to the study of phase transition in superconductors in 
the presence of magnetic fields.
\begin{figure}[t]
\center{{\includegraphics[height=110mm]{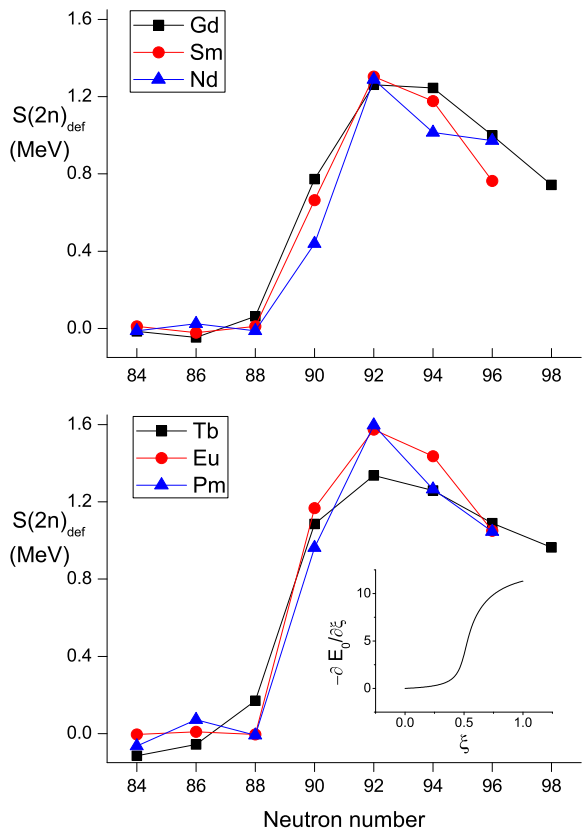}}}
\caption{
(color online). 
The contribution of deformation to the two-neutron separation
energies, $S(2n)_{\mathrm{def}}$ for even-even 
$_{60}$Nd-$_{62}$Sm-$_{64}$Gd nuclei (top) 
and odd-even $_{61}$Pm-$_{63}$Eu-$_{65}$Tb nuclei (bottom), plotted 
as a function of neutron number. The contribution is enhanced in odd-even
nuclei by approximately $300$ keV (at neutron number 92). Also the rise
between neutron numbers 88 and 90 is sharper in odd-even nuclei than in
even-even nuclei. In the limit $N\rightarrow \infty $ (no finite size
scaling) the quantity $S(2n)_{\mathrm{def}}$ should be zero before the 
critical value and finite and large after that. The expected behaviour of 
$-\frac{\partial E_0}{\partial \xi}$ for the U(5)-SU(3) transition and 
$N=10$ is shown in the inset.}
\label{fig3}
\end{figure}

Finally, having computed the equilibrium values, one can compute the 
total energies $E_{i}\left( N;\beta _{e},\gamma _{e};\xi ,\chi \right)$ 
which for the special case discussed here are given by 
$E_{K}\left (N;\beta_{e};\xi ;\Gamma \right )$ and shown in 
Fig.~\ref{fig2}. One can see again by comparing $E_{K}
$ with the energy of the purely bosonic system, $E_{B}$, that there is an
effect especially close to the critical value, $\xi _{c}$. The effect is not
so much in the total energy $E_{K}$ (where it is of order $1/N$) but in the
derivative of the total energy with respect to the control parameter, 
$\frac{\partial E_{K}}{\partial \xi }$. 

An important property of atomic nuclei is that they provide experimental
evidence for shape QPTs, in particular, of the spherical to axially-deformed 
transition (U(5)-SU(3) symmetry)~\cite{iac3,casten11,iac11}. 
Three signatures have been used to experimentally verify the occurrence of 
shape phase transitions in nuclei: 
(a)~the behavior of the order parameter ($\beta_e$) as a function of the 
control parameter, measured through the B(E2) values proportional 
to~$\beta_{e}^2$; (b)~the behavior of the ground state energies, 
measured through the two-neutron separation energies, $S_{2n}$; 
and (c)~the behavior of 
the gap between the ground state and the first excited $0^{+}$ state. 
Here for conciseness we concentrate only on 
$S_{2n} = -[E_{0}(N + 1) - E_{0}(N)]$, 
which can be related to the derivative of the
ground state energy, $E_0$, with respect to the control parameter, 
$\frac{\partial E_{0}}{\partial \xi }$. $S_{2n}$ 
can be written as a smooth contribution linear in the boson number $N$, 
plus the contribution of the deformation~\cite{iac1,cakirli09}
\begin{equation}
S_{2n} = -A_{2n}-B_{2n}N+S(2n)_{\mathrm{def}}~.
\label{Eq:S2n}
\end{equation}
In order to emphasize the occurrence of the phase transition it is
convenient to plot the deformation contribution only, obtained from the data
by subtracting the linear dependence, as a function of $N$. In previous
studies of the purely bosonic part it has been shown that $N$ is
approximately proportional to the control parameter $\xi$~\cite{iac3}. The
experimental values of $S(2n)_{\mathrm{def}}$ are shown in the top part of 
Fig.~\ref{fig3} for even-even nuclei (purely bosonic) and in the 
bottom part for odd-even nuclei (bosonic plus one fermion). 
They are obtained from the 
data~\cite{datasheets} with 
$A_{2n}=-14.61,\,-15.82,\, -16.997$ MeV for Nd-Sm-Gd, respectively, 
and $B_{2n}=0.657$ MeV, and with $A_{2n}=-15.185,\,-16.37,\,-17.672$ MeV 
for Pm-Eu-Tb, and $B_{2n}=0.670$ MeV. Precursors of the phase transition are
visible in all six nuclei between neutron numbers 88 and 90 in both, 
and, most importantly, appears to be enhanced in odd-even nuclei relative 
to the even-even case.

In conclusion, we have presented here a classical analysis of quantum phase
transitions in a system of $N$ bosons and one fermion (spin-boson system)
and shown that (i)~the addition of a fermion greatly modifies the critical 
value at which the phase transition occurs, and in some cases its nature; 
(ii)~the effect is similar to that of adding an external field; (iii)~there 
is experimental evidence for these phase transitions in odd-even nuclei at
neutron number 88-90. The effect of the odd fermion is about 20\% in 
$S(2n)_{\mathrm{def}}$. A quantal analysis, in which the Hamiltonian $H$ 
is diagonalized numerically for finite $N$, produces results similar to 
those of the classical analysis ~\cite{petrellis}. 
Our results are of interest not only for applications to
nuclei, but also for applications to other systems in which a fermion is
immersed in a bath of bosons, for example, the simple case of a spin 1/2
particle in a bath of harmonic oscillator bosons~\cite{lehur}.
Our analysis opens the way for a systematic study of QPTs in Bose-Fermi
systems, in particular, of shape phase transitions in odd-even nuclei. This
includes experimental studies and microscopic investigations using Density 
Functional Theory and/or other methods, in a way similar to what it has been 
done recently for the study of QPTs in purely boson systems 
(even-even nuclei)~\cite{casten11,vretenar09}.

This work was supported in part under DOE Grant No. DE-FG-02-91ER-40608 
and in part by a grant from the U.S.-Israel Binational Science Foundation.

\end{document}